\documentclass[prl,twocolumn,showpacs,keywords,preprintnumbers,amsmath,amssymb]{revtex4}
\usepackage{graphicx}
\usepackage{dcolumn}
\usepackage{bm}
\begin{document}
\title{The QCD jet-medium interactions}
\author{Shi-Yuan Li}
\email{lishy@sdu.edu.cn}
 \affiliation{Department of Physics,
Shandong University, Jinan, 250100, P. R. China}
\date{\today}

\begin{abstract}
In  relativistic heavy ion collisions, the interaction between the
 hard jet and the  quark-gluon matter
  has  an analogy of high energy scattering between hadrons. Hence
pionization   
 provides the key for   understanding  the  experimental results of
 heavy quark energy loss and   the
fragmentation function associate with jet.  Experimental tests for
this physical picture are suggested.
\end{abstract}

 \pacs{12.38.-t, 25.75.Bh, 13.85.Hd}

\keywords{Heavy quark energy loss, jet in A-A collision,
Bremsstrahlung, Pionization, Rapidity Region}

\maketitle

 One observation  on  confinement
is that it is  the property of the Quantum Chromodynamics (QCD)
vacuum (see e.g., \cite{tdleebook}). To explore this property, one
can investigate models based on  the QCD effective Lagrangian
\cite{delorenci}, as well as investigate the phenomenology of
(possible) phase transition from confinement to deconfinement. The
latter is the main purpose of Au-Au collision experiments on
relativistic heavy ion collider (RHIC) at BNL, and Pb-Pb collision
experiments  with higher energy as part of the project of large
hadron collider (LHC) at CERN (Here after, we refer them as A-A
collisions).

One of the genius ways to probe the state  after  the violent A-A
collision   is  ``jet tomography''  \cite{GW}, which relies on the
fact  that a  hard parton
 with large transverse momentum $p_T$  created  in the initial   collisions
interacts with the     quark-gluon matter (QGM,  this terminology
here refers to any possible state reached after the collision. It
can be hot, dense and deconfined,  or to the other extreme, only
ordinary  nucleus remnant) 
 and loses  energy. The energy loss can
reflect the information of the QGM. The   predicted  softer $p_T$
 spectra of hadrons 
relative to those in  hadron-hadron (h-h) collisions, and the
``mono-jet''
 have been  observed at RHIC, and play the
dominant r\^{o}le to estimate the properties, e.g., density, of the QGM,
via various formulations \cite{elgr}.
 These formulations  all
consider gluon bremsstrahlung induced by the coherent interactions
between the hard parton and QGM as the dominant partonic process
leading to energy loss, and the famous ``dead cone'' effect
 suppresses collinear gluon radiation from heavy quarks,
so that heavy quark energy loss could be dramatically weaker than
the light one \cite{dk}.
    However,  experiments \cite{HQEP} on $p_T$ spectra of
``non-photonic'' leptons from heavy quark (without distinguishing
charm or bottom) decays indicate that the energy loss of the heavy
quark is almost the same as that of the light
ones. 
Though  revisits on collisional  energy loss have narrowed
the gap between theories and data, there
is still necessity to study if some other
partonic process(es) has been ignored which
competes or even exceeds the above mentioned and  has
different dependence on quark mass.

Recently, the preliminary results in A-A collision by STAR
Collaboration \cite{Salur:2008hs,Putschke:2008wn}
 on the reconstruction
of  jets as well as associate fragmentation function  (FF) are
reported. Contrary to  expectations \cite{elgr}, no apparent
modification of the associate FF
 (relative to h-h collision)   is
observed at all.

In this paper we  revisit  ``pionization'' \cite{cw, pio}
common to all  high energy hadronic collisions,   
and apply this property to the ``effective
collision'' between the   hard jet and  QGM.
We argue that pionization  competes with
 coherent bremsstrahlung, 
eventually  exceeds the latter at enough high jet energy. The
{\it stochastic} (rather that collinear) nature of the emission angle
of the pionization products
 is helpful to understand   why the dead cone affects little and why the
FF associate with jet in A-A collision is
similar as that in h-h collisions.
  The corresponding process   for pionization in the Quantum Electrodynamics (QED)
case is  (fermion) pair production \cite{cw, pio}.
Worthy to be addressed is the fact that,  electromagnetic energy
loss of electron or muon in medium
 from ``pionization'' (electron-positron pair production)
exceeds that from photon  bremsstrahlung
at enough high energies   
  (see, e.g., \cite{slacr,cernr}).

  In the original study \cite{cw, pio},
pionization products correspond to the
 particles which constitute the ``rapidity plateau'' in
high energy hadronic scattering, and the rapidity axis is the beam
direction.  
 They may be important component of the QGM in A-A collision
 but are  irrelevant   to  the high $p_T$    
 partons created in the rare hard interactions, which is
employed as the hard probe.   
 What we investigate in this paper is the pionization in the {\it
effective high energy jet-QGM scattering},
which mimics the interactions between the hard probe parton and the
QGM. This  physical picture
 hence should be first clarified and established now. Jet
is a bundle of nearly collinear,  on shell particles, with the
invariant mass $M$ of them (as whole) much smaller than the
energy/virtuality $Q$ at which the jet (rigorously to say,  the
hard parton initializing the jet)
is created. We use $E$, $p_T$ to denote its energy and transverse
momentum, respectively, with $E \sim p_T \sim Q$ in our discussions.
 This energetic hard parton  evolutes from the space-time scale
 $\sim 1/p_T$ to   $\sim 1/M$ ($M<<p_T$)
by gluon radiation \cite{bk1},  and    the jet is preliminarily
shaped, defined by various jet algorithms at partonic level
(referred as ``preliminary jet'' in the following).
 Because  $1/M<<1fm$, just
the asymptotically   free region of QCD,
  any extra interaction of the hard parton with the remnant of the
hadrons (or nuclei) via momentum transfer  $\gtrsim M$  will lead
to extra suppression by (small) $\alpha_s$  and  (large)
denominator of the propagator. So the evolution during $1/p_T$ to
$1/M$  (hence the preliminary jet)
is hardly different whether it is created in h-h, A-A,
 $e^+$-$e^-$, or other collisions.
 However, the subsequent
evolutions to larger space-time scale will recognize the
``environment'' and depend on the concrete   scattering
processes. The uniqueness of ``central'' A-A collision
is the existence of the QGM 
rather than the vacuum in  other ``more simple'' scatterings.
 Because $M<<p_T \sim E$, {\it the jet as a whole  can be taken
as an  energetic composite particle with energy $E$ and mass $M$
hitting and passing through the
 QGM as target (Such is the jet tomography).}
 Each member (parton) of the jet, as the r\^{o}le of constituent,
  will interact with the QGM.
So {\it the jet interacting with the QGM is quite   in common as
the high energy hadronic or nuclear scatterings such as a proton
or nucleus hitting a target.} Based on such a physical picture,
without referring to any of the microscopic details of the QGM and
the energy and $p_T$ distribution of the preliminary jet, but
employing the  properties common to any hadronic collisions drawn
from experimental facts, one can qualitatively explain the above
mentioned  experimental ``paradoxes''  at RHIC \cite{HQEP,
Putschke:2008wn}. This is desirable since  QGM is ``uncharted'',
while the distribution of the preliminary jet is not predictable
if factorization is broken, 
at the case that the multiplicity is triggered hence  the
unitarity of the summation of soft interactions is violated.


Here we  clarify two cases of `induced radiations'.  The first one
is that induced by hard interactions.  For this case, a hard
process with a typical large momentum transfer  is triggered. The
Sudakov approximation, typically the double logarithm behaviour,
is the key property \cite{sudakov}. This describes the developing
of the jet, with the interactions between the jet and the remnants
(or other jets), or the initial radiation before the 
triggered hard interaction \cite{bk1}. 
In both cases, the radiation is collinear to the initial/final hard parton.
The space-time picture of these
processes is that, two issues taking part in  the hard interaction,
will never meet in space time after the hard collision (e.g., the
quark and anti-quark from $Z^0$ decay). If we choose the
coordinate of the hard interaction as the start point of the world
lines of these two stuffs,  these two lines will never meet again
in the future. This kind of induced radiation 
corresponds to that in any hard collisions except hard
probe in central heavy ion collision. For he former case 
 no energy residents in the mid-rapidity region (i.e., no QGM). For the
latter case, besides the induced radiation of 
the first case discussed above, there is the radiation induced 
by interaction between the
jet and the QGM,  without any hard trigger for this interaction
(The hard trigger for hard probe only denotes the creation of the
jet). If we draw the world line of the jet and the QGM, they will
meet again after the time of the jet creation. This means 
they will {\it collide}, after the time of the jet creation, 
as discussed in last paragraph.   The calculation of
this kind of process is under the approximation of high energy but
modest transverse (w.r.t. jet momentum) momentum \cite{cw}. 
The physical picture is as following.

In non-diffractive inelastic h-h collision, the ``energy loss'' of
the  initial colliding hadrons  is due to the the
multi-production, which ``uses'' and ``takes away'' part of  the
energy of them. This multi-particle system can be grouped into the
fireballs of the limiting fragmentation region \cite{zy} and the
pionization of the  central rapidity region  (we adopt the
definitions in \cite{cw, pio}). Even in the central A-A collision,
the structure of the multi-products is quite similar, as confirmed
by RHIC data (see, e.g., \cite{Nouicer:2002ks}).   These common
properties can be
applied to the effective jet-QGM scattering.   
The preliminary jet passes through the large-size and dense  QGM,
with multiple collisions,  so the probability
that all the collisions are elastic or diffractive
is vanishing.


 The rapidity distribution in the limiting fragmentation  region
  is approximately  similar for any
collision processes (from $e^+$-$e^-$ annihilation to central A-A
collision) \cite{Back:2006yw}.   
  We  extrapolate this   behaviour to the effective jet-QGM
scattering, i.e., assuming that the rapidity (w.r.t. the jet axis)
distribution of the limiting fragmentation of the preliminary jet
after hitting the QGM  is similar as those of  all the other
projectiles in various collisions. The reconstructed  jet   only
includes  the (nearly collinear) fragmentation  region.
So, neglecting the effect of transverse (w.r.t. the jet axis)
momentum distribution,
   the FF associate
with jet is hardly different whether in A-A or h-h collisions as seen at
RHIC \cite{Salur:2008hs,Putschke:2008wn}. 
  Further studies with identified particles
especially heavy hadrons ($D, B$) associate with jet will give
more clear results.

  The coherent gluon
bremsstrahlung induced by interacting with the QGM, is the
dominant contribution to the fragmentation, which  has been comprehensively
studied in literature \cite{elgr}. During its formation time
$k^0/k^2_T$,  the  radiated gluon of energy $k^0$
accumulates transverse momentum $k_T$,  
  Its angular distribution
concentrates at a  energy- and medium- dependent angle \cite{dk}.   
This is the microscopic basis of our extrapolating  the property of
limiting fragmentation to the fragmentation  of the preliminary jet.
 If the jet cone is  large enough,  the
``lost'' energy by coherent  bremsstrahlung can be fully
reconstructed into the jet energy.
So this mechanism can    
 contribute to the suppression of high $p_T$ hadron
but  not ``quenching'' of  the {\it whole jet} provided jet cone large enough.

The pionization   
is quite different from the limiting fragmentation.   
As is originally shown by Cheng and Wu \cite{pio}, for two
composite particle scattering, with limiting hypothesis and
Lorentz invariance, the inclusive distribution of poinization
products is definitely obtained to be $d\sigma/(dyd^2k_T) \sim
f({\bold k}_T)$, i.e., independent of $y$. Here $y$ and
${\bold k}_T$ are respectively  rapidity and transverse momentum
(Feynman got the similar result by argument based  on Lorentz
contraction at  high energy \cite{feynman}). Extrapolating this
common property to the high energy effective jet-QGM
scattering,  this way of energy deposition  from the
projectile jet is that the multiparticles
by pionization are stochastic in rapidity (w.r.t.
  the jet axis).
Neglecting  the masses of the pionization products, one can
identify the rapidity with pseudo-rapidity $y \to
\eta=-(1/2) \ln (tg(\theta/2))$.
 Then  the  angular distribution is obtained to be \vskip -0.36cm
\begin{equation}
 \frac{d\sigma}{d\theta}=\frac{d\sigma}{dy} \times |\frac{dy}{d\theta}|
\sim \frac{sin(\theta/2)}{cos^3(\theta/2)},
\end{equation} \vskip -0.2cm%
which favours large polar angle. No matter the initial  quark
is heavy or light, most of the pionization products after the effective collision lie in the
angular region much larger than the small dead cone angle. 
 If pionization contributes dramatically to energy loss,
 the dead cone affects little, consistent with the RHIC experiments
\cite{HQEP}.
    The STAR measurements \cite{Salur:2008hs,Putschke:2008wn}
 properly recognize that the jet could be  significantly
 ``broadened'' and the  energy of the initial parton
 could be  less reconstructed  by missing the
large angle particles.
    So  investigation with jet algorithms less biasing collinear
combination, e.g., JADE \cite{Bartel:1986ua},  is interesting.  
  Further more, if  a rapidity plateau of particles
 is  reconstructed along
the jet,  with the background particles,
including the  QGM reactions, are properly subtracted,
the energy loss of the initial parton via pionization could
be recovered. This provides {\it a measurement on the contribution
from pionization}.


 The pionization products are  characteristic of the phase space dominated by $k^+
\sim k^-$, i.e., modest rapidity, in center of mass frame.
The collinear approximation for the phase space, $k^0>>k_T$
($k^+>>k^-$ or $k^->>k^+$) \cite{elgr},   only valid for the
fragmentation (bremsstrahlung),  misses almost all the
pionization products.  For incorporating   pionization, one must
take into account the ``full'' phase space (all the rapidity
region).
 Systematic calculations  
 and arguments to all orders,  with the gluon propagators reggeized, 
 result in the extended  eikonal formula (\cite{cw} and refs.
therein, which also give brief accounts for the relation with the
``pomeron''), which gives the S-matrix of the scattering process
in impact parameter space via the eikonal operator.
  The extended eikonal formula implies  a physical picture of
multi-production of high energy scattering. It is a stochastic
process in which quanta are created and annihilated in a random
way \cite{cdyo,cw}. Though the transverse degrees of freedom
distribute according to the specific dynamics and the structure of
colliding issues, the longitudinal distribution is just the
rapidity plateau, with width proportional to $\ln s$,  and $s$ the
center of mass energy squared of the scattering.  This is the
microscopic basis of our extrapolating  the property of
pionization to the effective jet-QGM collision and indicates the
increase of the relative rate of the pionization energy loss  with the jet
energy due to the logarithmically   increasing
rapidity plateau  width.
 The pionization energy loss of the  preliminary jet can be estimated
to be
 \vskip -0.36cm
\begin{eqnarray}
<\Delta E> &= &\int_{y_{min}}^{y_{max}} dy d^2 k_T  \Big(\frac{d\sigma}
{dy d^2k_T}\Big ) k^0(y, k_T)\Big /\sigma_{incl} \nonumber \\
&=&\int_{y_{min}}^{y_{max}} dy d^2 k_T f({\bold k}_T) \frac{m_T}{2}
(e^y+e^{-y}) \Big / \sigma_{incl} \nonumber \\
&\sim & C
\int_{y_{min}}^{y_{max}} dy (e^y+e^{-y}).
\label{ome}
\end{eqnarray} \vskip -0.2cm
Here C comes from integration on  the transverse distribution,
including the information of  the concrete dynamics and structure
of the QGM, but similar for different kinds of energetic quark
jets, once the QGM fixed.   
  For the case that pionization is dominant,  this calculation  can also
include the contribution from the limiting fragmentation,
by a slightly modified value of $y_{min}$, $y_{max}$, based on the
mean value theorem of  integration.
  For central A-A collision in the laboratory frame, neglecting the asymmetry of its
thermal movements  and assuming the (left-right) symmetry of its  longitudinal
expansion, in average the QGM  can be taken as at rest. 
 So the rapidity $y$  of the created quanta    
 can take values from $y_{min} \sim 0$ to $y_{max}\sim A \ln(E/M)$.
 Here $E, M$, are respectively energy and mass of
the preliminary jet and $A$ is a constant.   $A$ can depend
on the dynamics, structure and size of the QGM.
 We then  conclude, without concrete values of the constant $C$ and $A$:

1)   $<\Delta E>/E \simeq C E^{A-1}/M^A$. This   power behaviour
of the  dependence on preliminary jet energy   relies on the
concrete width of the rapidity plateau (depending on the state of
QGM), and can   rise ($A >1$) or fall ($A<1$ ), comparing to the
LPM behaviour $\sim 1/E$ \cite{HQEP}.

2)    The ratio between  energy losses of two kinds of jets with same
energy $E$ but respective average  mass $M_1, ~M_2$ is
$r_{12}(E)=<\Delta E>_1/<\Delta E_2>\simeq (M_2/M_1)^A$.
  The details of the QGM (density,
 temperature, size, etc.) cancels.
   So this ratio  measured from RHIC in a range of jet energy, are
almost the same as those will be measured from LHC for the same
range of energy.
On the contrary, if  coherent bremsstrahlung is dominant, one gets
$r_{cb}(E) \sim (1-\frac{\hat q L^3}{2 E^2} M^2_c)/(1-\frac{\hat q
L^3}{2 E^2} M^2_b)$,
   by integrating  Eq. (16) in \cite{dk}, from 0
to maximum allowed energy, keeping only leading $\omega^{3/2}$
terms for large initial quark energy $E$.  For this  case,
the dependence on the transport coefficient (hence the density)
$\hat{q}$ and the size ($L$)  of the QGM enters,  and
  hot, dense, and large QGM can enhance the  differences
between different kinds of quarks.
The above two behaviours  are promising tests
of the physical picture we propose in this paper.   

3)  $M_1,~M_2$ are average masses of the preliminary jets
(``dressed parton'') rather than those of the
partons initializing the jets.  
 The initial
quark mass can introduce modifications to the  average jet mass,
  but the
difference is dramatically reduced by  dressing  quark mass to be
jet mass \footnote{By investigating  the average jet mass in 2-jet
events in $e^+$-$e^-$ annihilation for various initial quarks and
various center of mass energies, employing Durham algorithm in
Pythia/Jetset, we find that the proportion of average masses for
light, charm and bottom jets, $M_{jl}:M_{jc}:M_{jb}$,  varies from
$1:1.01:1.26$ to $1:1.01:1.08$,  for jet energies from $6GeV$ to
$20GeV$.}.
 This is exactly what RHIC data \cite{HQEP} indicate.
Furthermore,  the mass of light quark  jet can always be an
infrared-safe hard scale for perturbative QCD, while the light
quark mass ($\lesssim \Lambda_{QCD}$) can not.

A more feasible and decisive test of the above  physical picture
(though  prediction here only qualitative because of reasons in
the following), is the open charm meson (or more practical,
non-photonic lepton)
  nuclear modification factor $R_{AA}$ in the rapidity
interval $y = 1.2 \sim 2.2$ for central heavy ion collision at
RHIC or LHC. From the above discussion, especially point 1),
combining with the bremsstrahlung energy loss, at a linear
approximation of the dependence of $\Delta E$ on $E$, we get the
same expression for energy loss as in QED: $\Delta E=\alpha+\beta
E$. This means the larger of the jet energy, the more energy lost.
Hence, one can predict  more suppressed transverse momentum
spectrum hence  smaller $R_{AA}$ at $y = 1.2 \sim 2.2$ than the
value at around $y=0$, because for a definite $p_T$, one has larger 
total energy at $y = 1.2 \sim 2.2$ than at $y=0$. 
This is in fact implied by the $J/\Psi$
spectrum (though now mostly taken as indication of regeneration)
\cite{phejpsi}.

%
We have  confidence to suggest such a test since 
we have employed  the combination model  to reproduce the
$J/\Psi$ spectrum, by the same charm quark spectrum  to give the
open charm meson data at around $y=0$ \cite{shiyuan2}.
In such a uniform picture, it is easy  to fit one of the sectors, to predict the other. 
So one can extract from softer $J/\Psi$ spectrum at $y=1.2 \sim
2.2$, then predict a softer open charm (non-photonic lepton)
spectrum. The difficulty here is that, fitting an open spectrum to
$p_T$, can  almost predict $J/\Psi$ to $2 p_T$, on the contrary,
fitting the $J/\Psi$ to $P_T$, can only predict open charm to
$p_T/2$. The reason is that open charm is produced by the charm quark
'catching' a soft parton from QGM \cite{shiyuan1}, but $J/\Psi$ is
produced by combination of 2 parallel co-moving charm quarks. 
 So the $y=1.2 \sim 2.2$ results of $J/\Psi$ now \cite{phejpsi} can not give reliable
quantitative open charm prediction. But 
 the qualitative property is very clear, and this measurement is
very straightforward.

So far, to accommodate both the little modified FF associate with
jet as well as dramatic suppression of both heavy and light
particles of large $p_T$, that pionization  dominantly contributes
to energy loss  seems the most logical conclusion. 
  The key point for further precise calculations lies in the
factorization theorem for the multiplicity-associate observables, which we will
discuss elsewhere.

\vskip 0.3cm

{\large Acknowledgment}

The author would like to express his deep thanks to Prof. T. T.
Wu,  for  introducing the  book \cite{cw} greatly benefitting this
work,  and his inspiring discussions though not directly relevant
with this topic. The author also thanks the hospitality of PH-TH,
CERN, where early stage of this work was done. 
This work is partially supported by NSFC with grant No.
10775090, 10935012, and Natural Science Foundation of Shandong Province, China,  with grant No. ZR2009AM001.
Part of this work was done when the author visited UNIFEI, Brasil, supported by FAPEMIG. 
The hospitalities of ICE, UNIFEI, and the De Lorenci  family are greatly thanked.

\end{document}